\documentclass[sigconf, nonacm]{acmart}

\usepackage{pvldb}

\usepackage{enumitem}

\renewcommand\vldbdoi{XX.XX/XXX.XX}
\renewcommand\vldbpages{XXX-XXX}

\renewcommand\vldbavailabilityurl{URL_TO_YOUR_ARTIFACTS}

\begin{document}
\title{Crowd-OM: Crowdsourcing for Ontology Matching Validation}

\author{Zhangcheng Qiang}
\orcid{0000-0001-5977-6506}
\affiliation{
  \institution{Australian National University}
  \city{Canberra}
  \state{ACT}
  \country{Australia}
}
\email{qzc438@gmail.com}

\author{Weiqing Wang}
\orcid{0000-0002-9578-819X}
\affiliation{
  \institution{Monash University}
  \city{Melbourne}
  \state{ACT}
  \country{Australia}
}
\email{teresa.wang@monash.edu}

\author{Kerry Taylor}
\orcid{0000-0003-2447-1088}
\affiliation{
  \institution{Australian National University}
  \city{Canberra}
  \state{ACT}
  \country{Australia}
}
\email{kerry.taylor@anu.edu.au}

\begin{abstract}
Recent advances in large language models (LLMs) pose new challenges for ontology matching (OM). While OM systems built on LLMs have shown remarkable capabilities in discovering more matching candidates, traditional OM validation that relies on domain experts has become overwhelming. Although crowdsourcing can expose OM validation to large online communities, bias and errors from diverse annotators can reduce the quality of crowdsourcing for OM validation. In this study, we explore the use of crowdsourcing for OM validation and introduce a novel quality assurance crowdsourcing system called Crowd-OM.  In order to ensure the quality of crowdsourcing for OM validation, we propose three domain-specific mechanisms, namely differential trustworthiness, coherence pre-filling, and time-dependent modelling. Crowd-OM can be integrated with existing OM systems to enable human-in-the-loop validation. The evaluation shows its effectiveness in handling various annotators and different annotation scenarios. We discuss two real-world use cases and current limitations for improvement.
\end{abstract}

\maketitle

\vldbtopmatter

\section{Introduction}
\label{sec: introduction}

Ontology matching (OM) has been used in many real-world applications to enable semantic interoperability for knowledge graphs. Matching validation is a crucial step to ensure that the matching output aligns with real-world facts. This procedure is typically conducted by domain experts in surveys, interviews, and workshops. With the popularity of large language models (LLMs), OM systems have been widely incorporated with LLMs and LLM agents~\cite{qiang2023agent,hertling2023olala,giglou2024llms4om,zhang2024large,taboada2025ontology,nguyen2025kroma,sousa2025complex,song2026genom}. While such systems have shown advanced capabilities, rapid evaluation of LLMs poses several challenges for OM validation. In the Ontology Alignment Evaluation Initiative (OAEI)~\cite{oaei} campaign over the years, several annual reports find that LLM-based OM systems can discover more matching candidates, but these mappings are not present in current ground-truth references. Miscounting these disputed mappings has skewed the measurement of system performance. To tackle this issue, one popular approach is to employ harness engineering (e.g., chain-of-thoughts~\cite{wei2023chainofthought} and agent debate~\cite{du2024improving}) for LLMs, but LLM hallucinations are not easy to mitigate and can remain unavoidable~\cite{kalai2026evaluating}. For disputed mappings that require additional human validation, it is not easy to find domain experts and they may also have unavoidable human bias~\cite{tversky1974judgment}.

Crowdsourcing is particularly suitable for this scenario. Rather than conducting the validation within a group of domain experts, exposing this procedure to the large online community could significantly mitigate LLM hallucinations and reduce the negative effects of human bias. We consider crowdsourced participants who undertake tasks to be \textit{annotators}. Independent and diverse annotators exposing different options has proven to be an effective way to obtain less biased decisions~\cite{surowiecki2004wisdom}. However, crowdsourcing has several limitations. For example, the diversity of online annotators can include non-experts for validation, and their decisions are typically treated as equivalent to those of experts. This can lead to a phenomenon in which non-experts may dominate decision-making~\cite{selgin1996salvaging}. Moreover, crowdsourcing is labour-intensive and errors increase significantly when tasks require extensive mental effort~\cite{sweller1988cognitive}.

In this study, we explore the use of crowdsourcing for OM validation and introduce a novel crowdsourcing system. The system can be integrated with existing OM systems to enable crowdsourced OM validation. To tackle known issues in crowdsourcing, we propose three domain-specific mechanisms to improve the quality of crowdsourced OM validation. Our main contributions include:

\begin{itemize}[wide, noitemsep, topsep=0pt, labelindent=0pt]
\item We introduce a crowdsourcing system called Crowd-OM for OM validation. The system can be easily integrated with existing OM systems to enable interactive human-in-the-loop matching validation for OM tasks.
\item We propose three domain-specific mechanisms to improve the quality of crowdsourcing for OM validation.
\begin{itemize}[wide, noitemsep, topsep=0pt, labelindent=0pt]
\item Differential trustworthiness uses a set of seed pairs to assess each annotator's trustworthiness and apply the results to the overall agreement measure, therefore giving more weight to the expert's decision and reducing the weight of the non-experts' opinions.
\item Coherence pre-filling uses an in-built reasoner to pre-fill annotation pairs decisions from the annotator's completed answers, therefore reducing errors from manual input.
\item Time-dependent modelling uses the continuous editing mode to ensure the validation output is always up to date. When annotators change their decisions, the system tracks these changes and modifies accordingly in a timely manner.
\end{itemize}
\item We implement Crowd-OM with the three quality-assurance mechanisms on an existing OM system. We evaluate our proposed mechanisms in three controlled scenarios and demonstrate their effectiveness in improving the quality of crowdsourced OM validation.
\end{itemize}

The rest of the paper is organised as follows. Section~\ref{sec: related-work} reviews related work. Section~\ref{sec: formulation} presents the problem formulation. Section~\ref{sec: design} shows the system design of Crowd-OM. Section~\ref{sec: demonstration} demonstrates the use of Crowd-OM to extend existing systems for crowdsourced OM validation. Section~\ref{sec: evaluation} evaluates performance in different simulated scenarios. Section~\ref{sec: use-cases} describes two use cases, and Section~\ref{sec: limitations} discusses current limitations. Section~\ref{sec: conclusions} concludes the paper.

\section{Related Work}
\label{sec: related-work}

OM with human-in-the-loop (a.k.a. interactive OM) aims to overcome the bottleneck in automatic OM approaches that have reached the performance upper bound. Involving domain experts in OM validation has proven an effective way to utilise user feedback to improve OM performance~\cite{paulheim2013towards,dragisic2016user,li2019user}. In the era of LLMs, human-in-the-loop is integral. State-of-the-art LLM fine-tuning methods (e.g. RLHF~\cite{ouyang2022training} and DPO~\cite{rafailov2023direct}) leverage positive and negative user feedback to adjust LLMs for domain-specific tasks. Foreseeing the benefits of human-in-the-loop, the quality of user feedback becomes externally important. If the user's feedback does not provide sufficient information, LLMs may have very limited knowledge to learn or may even learn incorrect knowledge. Therefore, the annotators need to be carefully selected, and the answers need to be carefully filtered and processed.

Crowdsourcing is a popular technique in human-in-the-loop evaluation and is being used in a wide range of ontology engineering tasks. Traditionally, in the context of OM, validation is conducted by domain experts to ensure that the provided ground-truth reference alignments are correct for benchmarking. This approach requires extensive expert labour, which is infeasible for large-scale OM tasks. CROWDMAP~\cite{sarasua2012crowdmap} is an early work that proposes using a workflow model for crowdsourced OM. However, this method is highly dependent on the CrowdFlow platform, and the validation process provides limited support for resolving conflicting or inconsistent cases. To the best of our knowledge, this work is the first attempt to leverage crowdsourcing for quality OM validation with considerations of reliability and feasibility.

\section{Problem Formulation}
\label{sec: formulation}

Given the reference alignment $R$ and the matcher alignment $A$, OM validation considers non-disputed mappings to be the mappings that appear both in $R$ and $A$ (i.e. $A\cap R$). The mappings that exist only in $R$ (i.e. $R-A$) or only in $A$ (i.e. $A-R$) are disputed mappings and require additional human validation.

\section{System Design}
\label{sec: design}

Crowd-OM operates as a two-sided marketplace, similar to Airtasker~\cite{airtasker}. A group of system users can post their annotation tasks, and another group of system users can participate in them. We initialise three different types of system users.

\begin{enumerate}[wide, noitemsep, topsep=0pt, labelindent=0pt]
\item Administrator: The administrator is the superuser of the system. The administrator is responsible for uploading the original datasets and controlling the open/close status. The administrator can also view details of the system users, such as the annotation tasks in which they participate and the decisions they made on each annotation pair.
\item Developer: The developer is a specific type of system user. They are allowed to upload the alignment produced by the matcher into the system and to specify the customised settings for their annotation task. Developers are not allowed to participate in annotation tasks produced by their own matchers.
\item Annotator: The annotator is the main user of the system. They are allowed to annotate the same datasets multiple times for different tasks posted by different matchers.
\end{enumerate}

We allow a single user to have multiple roles in the system. For example, a developer can serve as an annotator by participating in annotation tasks posted by other matchers. We consider a developer to be a user who has registered at least one matcher in the system, whereas an annotator is a user who has started at least one annotation task. Annotators must complete registration (including reading the information sheet and signing the consent form) before participating in the annotation task for data collection.

\begin{figure}[htbp]
\centering
\includegraphics[width=0.9\linewidth]{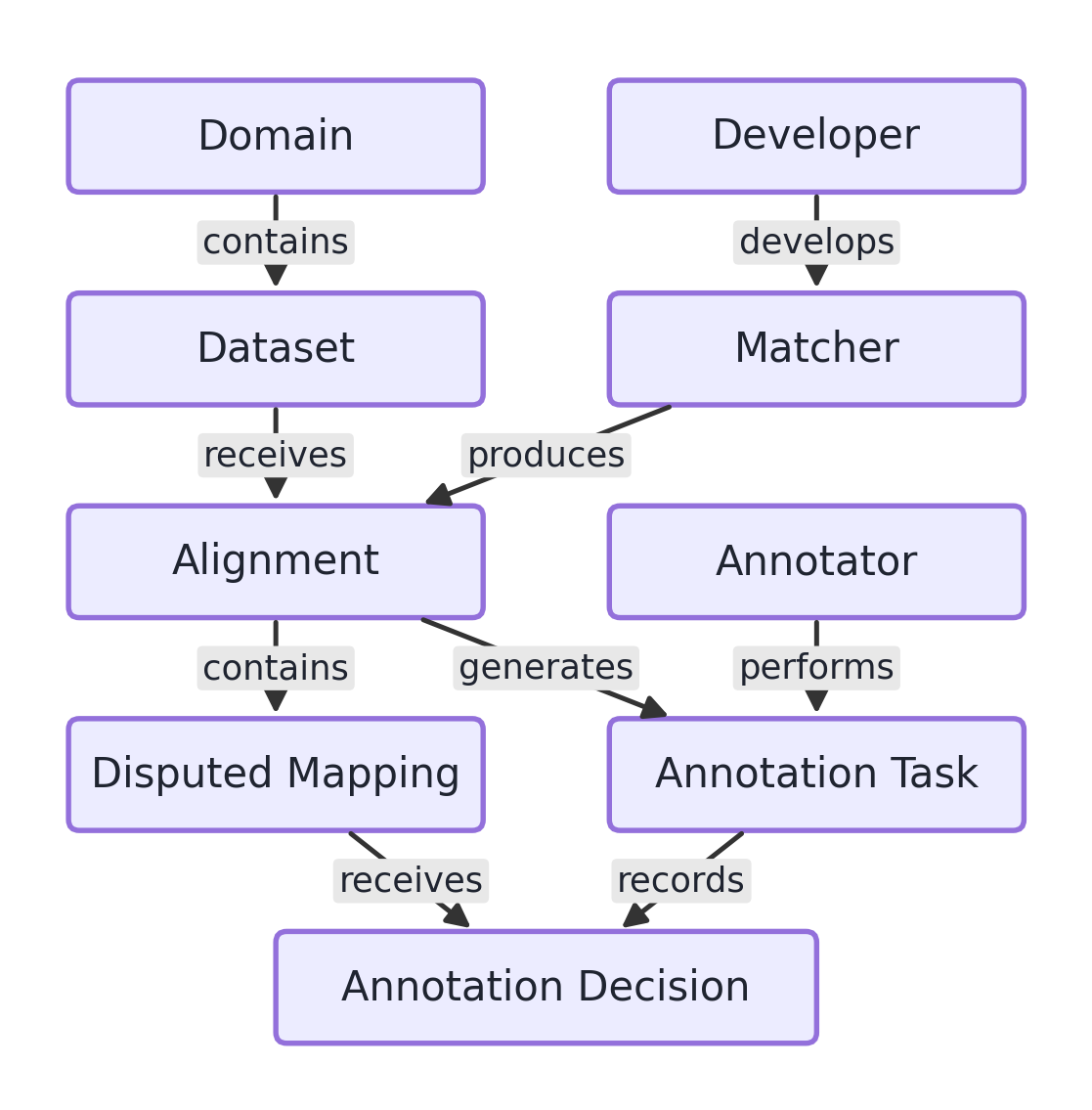}
\caption{Conceptual model of Crowd-OM.}
\Description{Conceptual model of Crowd-OM.}
\label{fig: conceptual-model}
\end{figure}

Figure~\ref{fig: conceptual-model} shows the conceptual model of Crowd-OM. The system consists of datasets from different domains and matchers from different developers. When a matcher produces a system alignment, the result is compared with the corresponding dataset to find disputed mappings. These mappings are formed into an annotation task. Annotators need to make decisions on each disputed mapping contained in the annotation task.

Each annotator needs to determine whether the annotation pair (i.e. two entities in the disputed mappings) are ``Equivalent'' or ``Not Equivalent''. They can view the class hierarchy, logic restriction, and textual description side by side to facilitate their decision on equivalence. Crowd-OM follows the majority voting rule. A disputed mapping is considered equivalent if more than 50\% of annotators vote ``Equivalence''. Some domains (e.g. biomedical) require high precision; the system allows the confidence threshold to be adjusted from 50\% to 100\%. To ensure the quality of crowdsourcing for OM validation, we propose three novel quality-assurance mechanisms into the system.

\subsection{Differential Trustworthiness}

Crowdsourcing does not restrict non-experts from participating in annotation tasks. The number of non-experts is often greater than that of experts, so the votes of non-experts can dominate the votes of experts. We expect the experts' decisions to carry more weight.

We construct a set of seed pairs to complement the original base pairs. The seed pairs are used to examine the expertise of annotators, and therefore determine the trustworthiness of their decisions. These seed pairs have clear answers of ``Equivalent'' or ``Not Equivalent''. The seed pairs need to be mixed with the original annotation pairs to avoid recognition. We use the following logic to generate positive and negative seed pairs.

\begin{itemize}[wide, noitemsep, topsep=0pt, labelindent=0pt]
\item Positive seed pairs are generated from the non-disputed mappings and consist of two parts: (1) Trivial: entity names (or labels) are similar. (2) Non-trivial: entity names (or labels) are different.
\item Negative seed pairs are generated using the following logic: (1) Trivial: If $(e1, e2)$ exists in non-disputed mappings, then $(e3, e2)$ where $e3 \equiv e1$ should not be correct. This assumes these mappings are one-to-one mappings. (2) Non-trivial: $(e1,e2)$ exist in disputed mappings and fail to pass the ontology coherence check.
\item To ensure that the difficulty of positive and negative seed pairs is similar, we compute the total number and the distribution of trivial and non-trivial rates for positive seed pairs and apply the same ratio to generate negative seed pairs.
\end{itemize}

Given the number of seed pairs $m$ and the number of correct answers $n$ given by the annotator $u$, the trustworthiness $\tau$ is defined as $\tau(u) = n/m$. When moving to the original annotation pairs (i.e. non-seed pairs), $\tau$ is used as a weight applied to each annotator's answer. We define $d(u,q)\in \{0,1\}$ as the score received from the annotator $u$ for an original annotation pair $q$, in which $0$ refers to the answer ``Not Equivalent'' and $1$ refers to the answer ``Equivalent''. The weighted agreement score accumulated for each $q$ is the product of each annotator's answer $d(u,q)$ and their trustworthiness $\tau$, divided by the number of annotators $N$. Hence, the weighted agreement score $w(q)\in (0,1)$ is computed as:

\begin{equation}
w(q) = \frac{1}{N} \sum_{u=1}^{N} d(u,q) \cdot \tau(u)
\end{equation}

\subsection{Coherence Pre-Filling}

Crowdsourcing expects annotators' attention and involvement to be consistent throughout the annotation session. However, several studies have shown that annotators' engagement inevitably declines with time-consuming tasks. We aim to reduce the workload for annotators by automatic pre-filling.

We use the coherence check to achieve this function. While each annotator's decision creates an assertion rule, these rules can be used to infer other supporting or conflicting annotation pairs. This can be applied both within the datasets and between datasets within the same domain. For example, within the dataset, if the annotator labelled $A \equiv B$, then $C \not\equiv B$ can be pre-filled in a one-to-one mapping task; Between datasets within the same domain, if the annotator labelled $A \equiv B$ in one dataset and $B \equiv C$ in another dataset, then a transitive equivalence $A \equiv C$ can be inferred and pre-filled if it appears in the dataset from the same domain.

A reasoner is used to automate this process. We construct an \textit{ontology network} and ask the reasoner to infer the explicit assertions as pre-fills. The \textit{ontology network} contains the original ontology as the base ontology and forms each annotation decision as an assertion within the ontology. For the coherence check between datasets within the same domain, we also combine multiple ontologies.

\subsection{Time-Dependent Modelling}

Typically, there is no persistent ground truth for OM tasks. The reference alignments generated by domain experts are based on their own opinions. For example, ``NCI'' commonly refers to ``National Cancer Institute'' globally, but within the context of Australian research, this abbreviation is more likely to refer to ``National Computational Infrastructure''. Our system allows annotators to modify their decisions over time and to assign a timestamp label to the validation output. When an annotator changes the decision, the overall crowdsourced decision is expected to change accordingly.

\begin{itemize}[wide, noitemsep, topsep=0pt, labelindent=0pt]
\item If the annotator changes the decision on the original annotation pairs, this action directly affects the crowdsourced decision.
\item If the annotator changes the decision on the seed pairs, this action may change their trustworthiness and further affect the crowdsourced decision on each original annotation pair.
\end{itemize}

\section{System Demonstration}
\label{sec: demonstration}

We demonstrate how Crowd-OM can extend existing OM systems to enable crowdsourced human-in-the-loop validation. We select Agent-OM~\cite{qiang2023agent} as the matcher, and the matching output is obtained from the OAEI 2025 campaign~\cite{qiang2025oaei}.

\begin{itemize}[wide, noitemsep, topsep=0pt, labelindent=0pt]

\item \textbf{Administrator Actions.} The system only allows the administrator to upload the reference alignment $R$. The administrator needs to assign the dataset to one registered domain. If the domain does not exist, the administrator can register a new domain. Each dataset requires three files to upload: the two ontology files and the reference alignment between them. For example, the administrator registers a domain called ``conference'' and adds two datasets ``cmt-conference'' and ``cmt-conof'' under the ``conference'' domain. The administrator can edit the settings of the datasets and domains by setting a confidence threshold and opening or closing the domain for annotation. For example, the administrator sets the 50\% confidence threshold for the ``conference'' domain and opens this domain for annotation.

\item \textbf{Developer Actions.} The system only allows the developer to upload the matcher alignment $A$. The developer needs to register the matcher and provide a short description. By doing so, they automatically become the developer of the matcher. The developer can also add other developers. Once the developer completes the upload, the system computes the difference between the matcher alignment $A$ and the reference alignment $R$ to find disputed mappings and form an annotation task. The developer can edit the settings for the annotation task. For example, the developer can enable or disable the differential trustworthiness and coherence pre-filling mechanisms as follows. (1) If differential trustworthiness is being activated, the system automatically generates a set of positive and negative seed pairs. These seed pairs labelled as ``seed'' are visible to developers but not to annotators. (2) If coherence pre-filling is being activated, the system performs a coherence check based on the inputs from the annotator and pre-fill annotation pairs. Annotation pairs labelled as ``pre-filling'' are visible to both developers and annotators, with a short explanation of the description logic involved.

\item \textbf{Annotator Actions}. The system allows the annotator to select an annotation task produced by a single matcher for a specific dataset. They can only annotate the tasks with the matcher for which they are not a registered developer, and the dataset under the domain is currently open for annotation. There are three options available for each annotation pair: ``Equivalent'', ``Not Equivalent'', and ``N/A''. (1) The trustworthiness is computed when the annotator completes the annotation task (i.e. there is no ``N/A'' annotation within the dataset). (2) The coherence pre-filling continuously monitors the annotator's decisions to identify new pre-fills. (3) The annotator can change the decision on each annotation pair at any time. The trustworthiness and coherence pre-filling will respond to changes in the decision accordingly.

\end{itemize}

\section{System Evaluation}
\label{sec: evaluation}

In this section, we simulate the diversity of the annotator groups and different system settings. The purpose is to demonstrate the system's effectiveness in handling real-world challenges of using crowdsourcing for OM validation, such as low expert ratio and knowledge level, or high rates of annotation errors.

We select six OAEI tracks comprising 45 datasets for evaluation. The matcher output is obtained from the Agent-OM results in the OAEI 2025 campaign. For each controlled environment, we simulate 100 human annotators in the system. We use the default confidence threshold of 50\% and set the seed number to 42 for reproducibility.

\begin{itemize}[wide, noitemsep, topsep=0pt, labelindent=0pt]

\item \textbf{Differential Trustworthiness.} We simulate 100 human annotators with a mixture of experts and non-experts. Experts are expected to have a trustworthiness greater than 0.5, while non-experts are expected to have a trustworthiness less than 0.5. The trustworthiness reflects the accuracy of the annotator's answers. For example, a trustworthiness of 0.7 means that the annotator has a 70\% probability of labelling the TP as ``Equivalent'' and a 30\% probability of labelling the TP as ``Not Equivalent''. Here, we assume that the reference alignments $R$ provided by the OAEI track are complete. The mappings that exist only in $R$ (i.e. $R-A$) are true-positive mappings (TPs), while the mappings only in $A$ (i.e. $A-R$) are false-positive mappings (FPs).

Figure~\ref{fig: simulation-dt-expert-ratio} compares the discovery rates of TPs and FPs as the expert ratio varies from 0.0 to 1.0. Although there is no difference when the expert radio is very low or very high, our differential trustworthiness mechanism shows significant improvements when the expert radio is in the range [0.1, 0.6], which is more common in real-world scenarios. Figure~\ref{fig: simulation-dt-expert-lower} compares the discovery rates of TPs and FPs as the expert knowledge lower bound varies from 0.5 to 1.0. We can see that our differential trustworthiness is particularly useful when the expert knowledge lower bound is less than 0.85. In reality, experts can also make mistakes and achieving the expert knowledge lower bound above 0.85 is almost impossible.

\begin{figure}[htbp]
\centering
\includegraphics[width=1\linewidth]{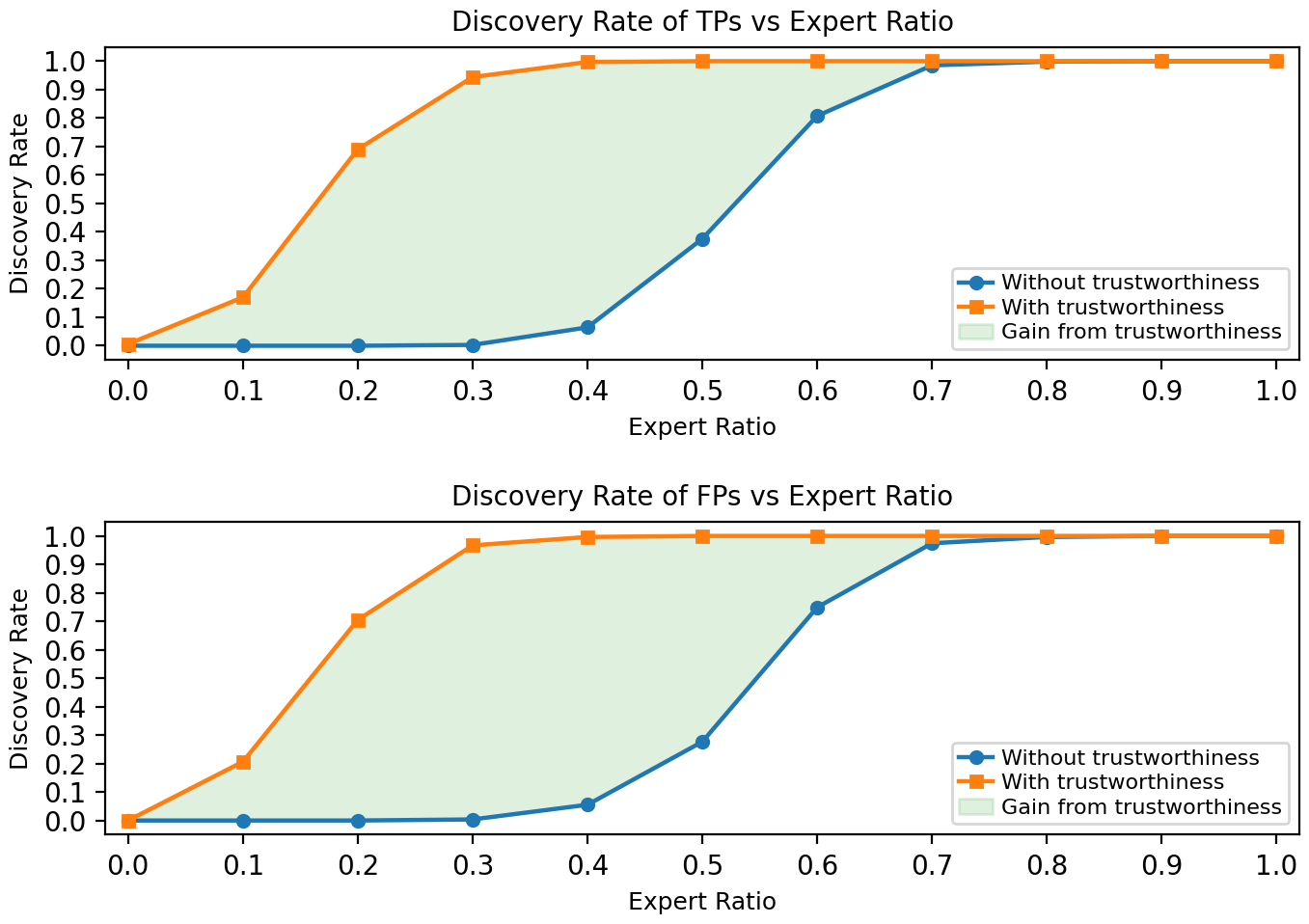}
\caption{Discovery rates of TPs and FPs vs expert ratio.}
\Description{Discovery rates of TPs and FPs vs expert ratio.}
\label{fig: simulation-dt-expert-ratio}
\end{figure}

\begin{figure}[htbp]
\centering
\includegraphics[width=1\linewidth]{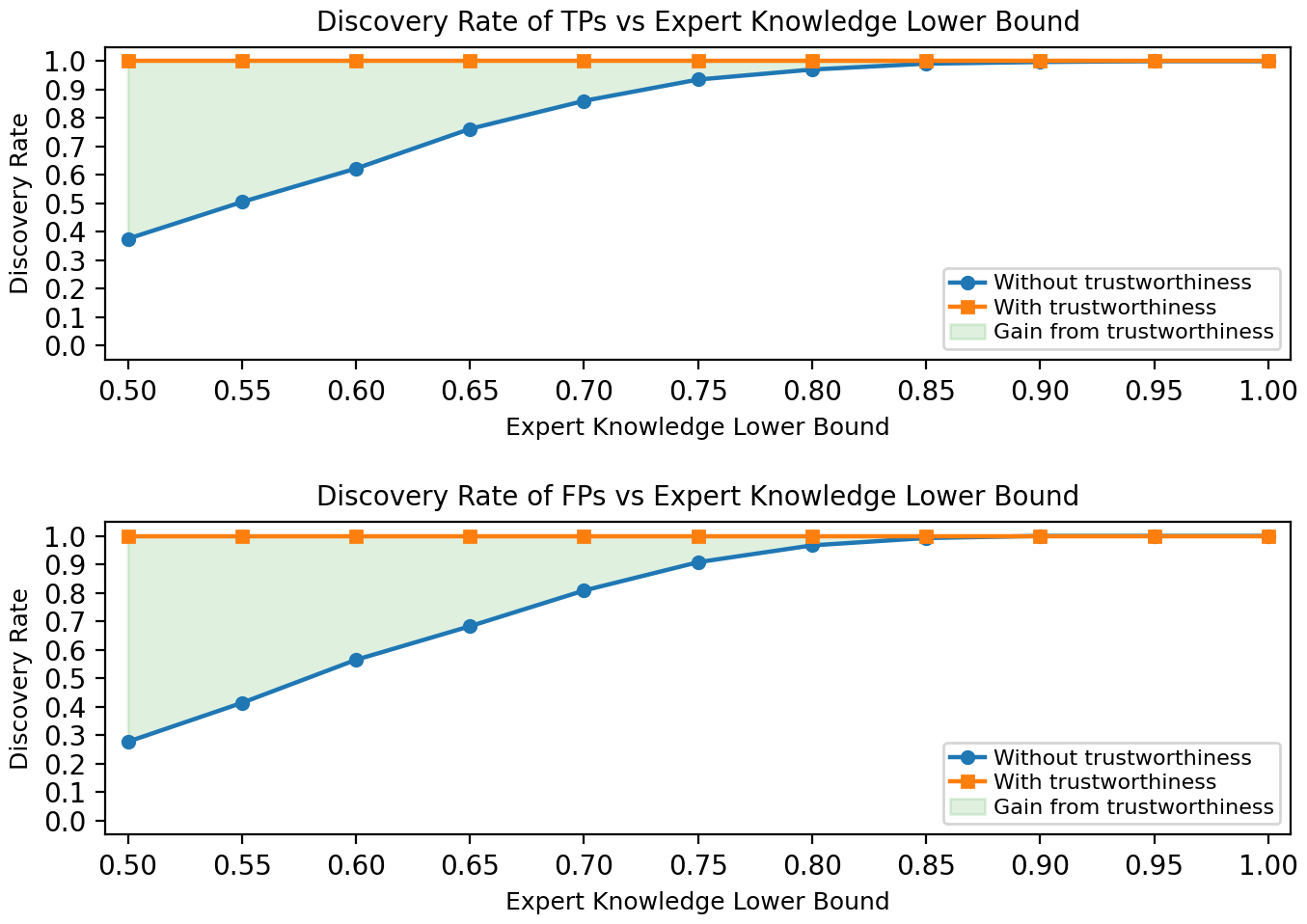}
\caption{Discovery rates of TPs and FPs vs expert knowledge.}
\Description{Discovery rates of TPs and FPs vs expert knowledge.}
\label{fig: simulation-dt-expert-lower}
\end{figure}

\item \textbf{Coherence Pre-Filling.} We simulate 100 human annotators and each annotator's decision is randomly selected from ``Equivalent'' or ``Not Equivalent''. We do not specify whether annotators are experts or non-experts. For the purpose of assessing the propagation behaviours of coherence pre-filling, we expect this function to work for all annotators. Here, we define an \textit{anchor ontology} for each domain and call datasets containing alignments for this ontology \textit{anchor datasets}. We only simulate the annotator's input on \textit{anchor datasets} and leave others for pre-filling candidates. We use \textit{coverage} to describe the fraction of annotated pairs in the \textit{anchor datasets}.

Figure~\ref{fig: simulation-cpf-type} compares the number of pre-fills by type as coverage varies from 0.1 to 1.0. Only three types exist in the OAEI datasets we evaluated: transitive equivalence, subsumption negativity, and disjointness. Overall, we can see that the coverage within [0.3, 0.8] yields more pre-fills. A low fraction of annotated pairs reduces the number of assertions to discover, whereas a high fraction of annotated pairs leaves little space for pre-filling to be functional. Figure~\ref{fig: simulation-cpf-domain} compares the number of pre-fills by domain as coverage varies from 0.1 to 1.0. We can see that the propagation behaviours of coherence pre-filling vary across different domains. This may be because the ontology design in each domain follows its own well-agreed guidelines, thereby restricting the use of logic expressions.

\begin{figure}[!t]
\centering
\includegraphics[width=1\linewidth]{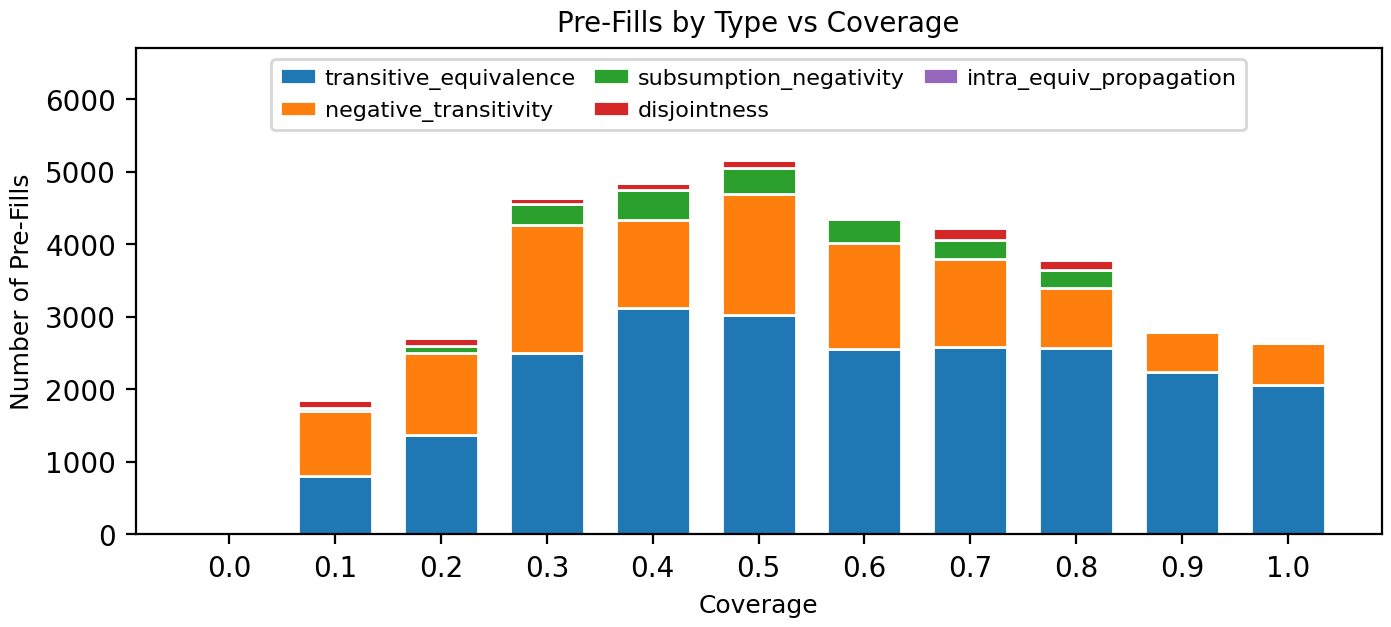}
\caption{Pre-fills by type vs coverage.}
\Description{Pre-fills by type vs coverage.}
\label{fig: simulation-cpf-type}
\end{figure}

\begin{figure}[!t]
\centering
\includegraphics[width=1\linewidth]{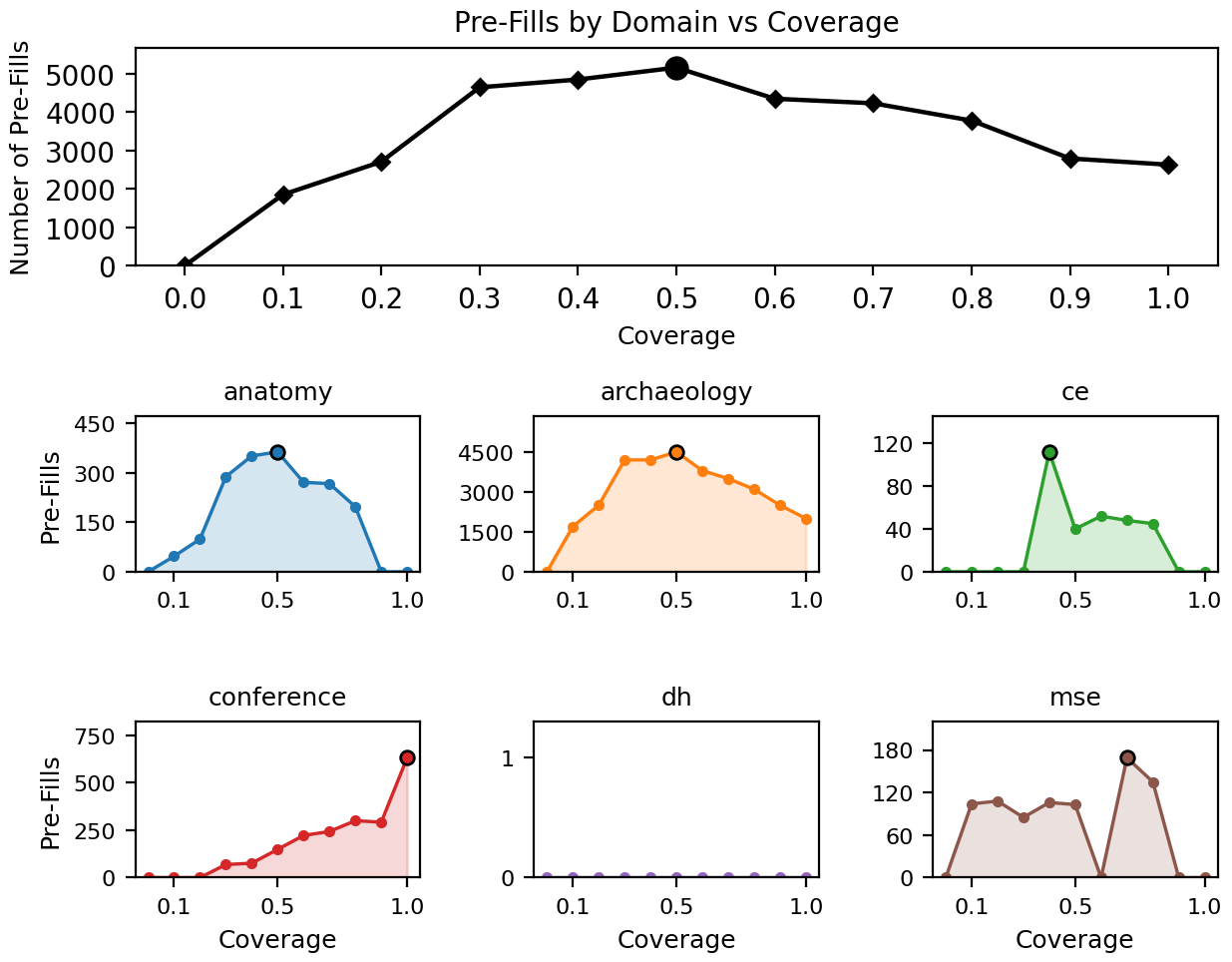}
\caption{Pre-fills by domain vs coverage.}
\Description{Pre-fills by domain vs coverage.}
\label{fig: simulation-cpf-domain}
\end{figure}

\item \textbf{Time-Dependent Modelling.} We simulate 100 human annotators (50\% experts and 50\% non-experts) with varying trustworthiness. Similarly, we assume that TPs are mappings that exist only in the reference and FPs are mappings that exist only in the matcher. Here, we define 0.5/0.5 as both non-experts and experts having trustworthiness 0.5, while 0.1/0.9 means non-experts with trustworthiness 0.1 and experts with trustworthiness 0.9.

Figures~\ref{fig: simulation-tdo-non-seed} and~\ref{fig: simulation-tdo-seed} compare the discovery rates of TPs and FPs as the correction rate varies from 0.0 to 1.0.  We can see that correction in non-seed pairs always increases the discovery rate and follows a linear incremental pattern. For correction in seed pairs, there is only a significant increase with a low spread ratio of 0.5/0.5 or 0.4/0.6 and a low correction fraction within [0.0, 0.2]. Note that the seed pair correction is a cascading operation, as the annotator will be given new correct non-seed pairs due to the increase in trustworthiness. The results indicate that our time-dependent modelling can capture individual decision changes over time with corresponding changes in the overall decision. It also works well with differential trustworthiness to fairly assess changes in the annotator's decisions.

\begin{figure}[htbp]
\centering
\includegraphics[width=1\linewidth]{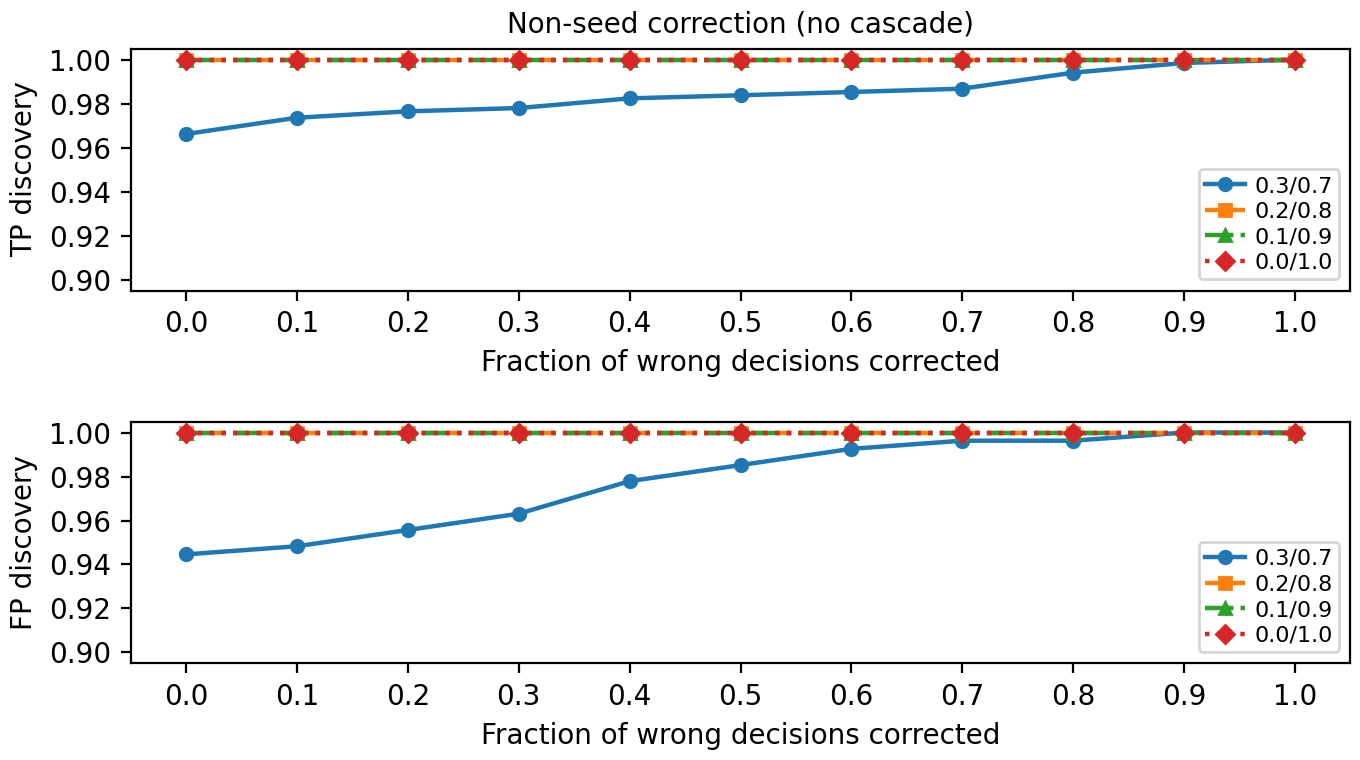}
\caption{Discovery rates of TPs and FPs vs fraction of wrong decisions corrected (non-seed pairs).}
\Description{Discovery rates of TPs and FPs vs fraction of wrong decisions corrected (non-seed pairs).}
\label{fig: simulation-tdo-non-seed}
\end{figure}

\begin{figure}[htbp]
\centering
\includegraphics[width=1\linewidth]{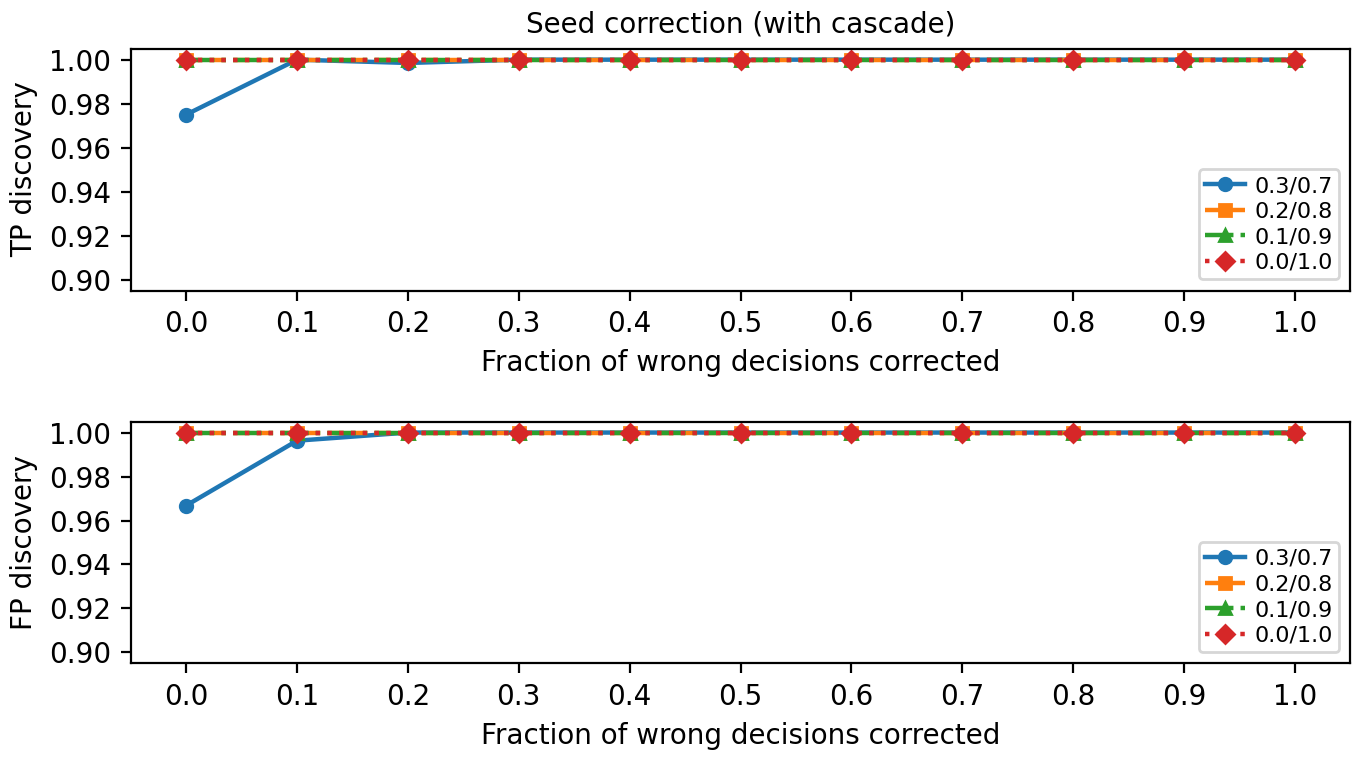}
\caption{Discovery rates of TPs and FPs vs fraction of wrong decisions corrected (seed pairs).}
\Description{Discovery rates of TPs and FPs vs fraction of wrong decisions corrected (seed pairs).}
\label{fig: simulation-tdo-seed}
\end{figure}

\end{itemize}

\section{Use Cases}
\label{sec: use-cases}

Having evaluated Crowd-OM for improving the quality of OM validation, we present two real-world use cases.

\begin{itemize}[wide, noitemsep, topsep=0pt, labelindent=0pt]

\item \textbf{Reference Evolution.} OAEI is one of the largest communities focused on OM. Over the years, the reference alignment used across several tracks has been out of date. Since the introduction of ChatGPT~\cite{chatgpt} in 2022, new matchers incorporating LLMs have identified more matching candidates that are not present in current references. These mappings seem to be correct, but they lack expert human validation to ensure their quality. While existing reference alignments are curated by domain experts, they may exhibit confirmation bias, in which their judgements are based on their preconceived notions. Crowd-OM can mitigate human bias in this case.

\item \textbf{Ontology Harmonisation.} In the building industry, ontology harmonisation has received significant attention due to the emerging need to integrate building information systems from different vendors. Despite the foundation established by the harmonisation efforts between Brick Schema~\cite{balaji2016brick}, RealEstateCore~\cite{hammar2019realestatecore}, and Project Haystack~\cite{john2020project}, achieving semantic interoperability between these ontologies remains a challenge. Many mappings in these ontologies commonly use abbreviations and domain-specific terminologies and require extensive domain expertise for validation. Crowd-OM can be used as an alternative to expert-based validation.

\end{itemize}

\section{Limitations}
\label{sec: limitations}

It is recommended to have a diverse set of participants for annotation tasks, but this is not controllable in crowdsourcing systems. We consider the bias caused by homogeneity to be minor when there is a considerable number of annotators. Moreover, real-world annotations can be more complicated than those in controlled simulations. For example, we assume the annotator's performance is consistent across seed and non-seed pairs, but it could be that the annotator performs very well on seed pairs but poorly on the real non-seed pairs. These edge cases need to be carefully identified and preprocessed before fitting to the models for analysis.

Crowd-OM currently focuses on the most common one-to-one equivalent matching, but it can be extended to more complex matching tasks. For example, the logic for generating equivalent seed pairs can be extended for subsumption: (1) Positive: If $A \equiv B$ and $C \subseteq A$, then $C \subseteq B$. (2) Negative: If $A \not\equiv B$ and $C \subseteq A$, then $C \not\subseteq B$. Coherence pre-filling can be modified to support subsumption tasks in similar ways.

\balance

\section{Conclusions}
\label{sec: conclusions}

In this paper, we propose a novel crowdsourcing system Crowd-OM for OM validation with three quality-assurance mechanisms. We present the system design and illustrate a system demonstration. The evaluation shows that the proposed quality-assurance mechanisms improve the system's robustness and overcome the inherent challenges of crowdsourcing for OM validation. In the future, we plan to adopt Crowd-OM in real-world use cases and contribute to the creation of robust benchmarks for OM tasks.

\begin{acks}
The authors thank the Commonwealth Scientific and Industrial Research Organisation (CSIRO) for supporting this project. The authors thank Jing Jiang from Australian National University (ANU) for providing insights on the experiment design. Weiqing Wang is the recipient of an Australian Research Council Discovery Early Career Researcher Award (project number \href{https://dataportal.arc.gov.au/NCGP/Web/Grant/Grant/DE250100032}{DE250100032}) funded by the Australian Government.
\end{acks}

\bibliographystyle{ACM-Reference-Format}
\bibliography{qiang-bibliography-vldb}


\begin{thebibliography}{28}


\ifx \showCODEN    \undefined \def \showCODEN     #1{\unskip}     \fi
\ifx \showDOI      \undefined \def \showDOI       #1{#1}\fi
\ifx \showISBNx    \undefined \def \showISBNx     #1{\unskip}     \fi
\ifx \showISBNxiii \undefined \def \showISBNxiii  #1{\unskip}     \fi
\ifx \showISSN     \undefined \def \showISSN      #1{\unskip}     \fi
\ifx \showLCCN     \undefined \def \showLCCN      #1{\unskip}     \fi
\ifx \shownote     \undefined \def \shownote      #1{#1}          \fi
\ifx \showarticletitle \undefined \def \showarticletitle #1{#1}   \fi
\ifx \showURL      \undefined \def \showURL       {\relax}        \fi
\providecommand\bibfield[2]{#2}
\providecommand\bibinfo[2]{#2}
\providecommand\natexlab[1]{#1}
\providecommand\showeprint[2][]{arXiv:#2}

\bibitem[\protect\citeauthoryear{Airtasker}{Airtasker}{nd}]%
        {airtasker}
\bibfield{author}{\bibinfo{person}{Airtasker}.} \bibinfo{year}{n.d.}\natexlab{}.
\newblock \bibinfo{title}{Airtasker}.
\newblock
\newblock
\urldef\tempurl%
\url{https://www.airtasker.com/}
\showURL{%
Retrieved May 1, 2026 from \tempurl}


\bibitem[\protect\citeauthoryear{Babaei~Giglou, D'Souza, Engel, and Auer}{Babaei~Giglou et~al\mbox{.}}{2024}]%
        {giglou2024llms4om}
\bibfield{author}{\bibinfo{person}{Hamed Babaei~Giglou}, \bibinfo{person}{Jennifer D'Souza}, \bibinfo{person}{Felix Engel}, {and} \bibinfo{person}{S{\"o}ren Auer}.} \bibinfo{year}{2024}\natexlab{}.
\newblock \showarticletitle{{LLMs4OM}: Matching Ontologies with Large Language Models}. In \bibinfo{booktitle}{\emph{The Semantic Web: {ESWC} 2024 Satellite Events}}. \bibinfo{publisher}{Springer}, \bibinfo{address}{Hersonissos, Crete, Greece}, \bibinfo{pages}{25--35}.
\newblock
\urldef\tempurl%
\url{https://doi.org/10.1007/978-3-031-78952-6_3}
\showDOI{\tempurl}


\bibitem[\protect\citeauthoryear{Balaji, Bhattacharya, Fierro, Gao, Gluck, Hong, Johansen, Koh, Ploennigs, Agarwal, Berges, Culler, Gupta, Kj\ae{}rgaard, Srivastava, and Whitehouse}{Balaji et~al\mbox{.}}{2016}]%
        {balaji2016brick}
\bibfield{author}{\bibinfo{person}{Bharathan Balaji}, \bibinfo{person}{Arka Bhattacharya}, \bibinfo{person}{Gabriel Fierro}, \bibinfo{person}{Jingkun Gao}, \bibinfo{person}{Joshua Gluck}, \bibinfo{person}{Dezhi Hong}, \bibinfo{person}{Aslak Johansen}, \bibinfo{person}{Jason Koh}, \bibinfo{person}{Joern Ploennigs}, \bibinfo{person}{Yuvraj Agarwal}, \bibinfo{person}{Mario Berges}, \bibinfo{person}{David Culler}, \bibinfo{person}{Rajesh Gupta}, \bibinfo{person}{Mikkel~Baun Kj\ae{}rgaard}, \bibinfo{person}{Mani Srivastava}, {and} \bibinfo{person}{Kamin Whitehouse}.} \bibinfo{year}{2016}\natexlab{}.
\newblock \showarticletitle{Brick: Towards a Unified Metadata Schema For Buildings}. In \bibinfo{booktitle}{\emph{Proceedings of the 3rd ACM International Conference on Systems for Energy-Efficient Built Environments}}. \bibinfo{publisher}{ACM}, \bibinfo{address}{Palo Alto, CA, USA}, \bibinfo{pages}{41--50}.
\newblock
\urldef\tempurl%
\url{https://doi.org/10.1145/2993422.2993577}
\showDOI{\tempurl}


\bibitem[\protect\citeauthoryear{Dragisic, Ivanova, Lambrix, Faria, Jim{\'e}nez-Ruiz, and Pesquita}{Dragisic et~al\mbox{.}}{2016}]%
        {dragisic2016user}
\bibfield{author}{\bibinfo{person}{Zlatan Dragisic}, \bibinfo{person}{Valentina Ivanova}, \bibinfo{person}{Patrick Lambrix}, \bibinfo{person}{Daniel Faria}, \bibinfo{person}{Ernesto Jim{\'e}nez-Ruiz}, {and} \bibinfo{person}{Catia Pesquita}.} \bibinfo{year}{2016}\natexlab{}.
\newblock \showarticletitle{User Validation in Ontology Alignment}. In \bibinfo{booktitle}{\emph{The Semantic Web -- ISWC 2016}}. \bibinfo{publisher}{Springer}, \bibinfo{address}{Kobe, Japan}, \bibinfo{pages}{200--217}.
\newblock
\urldef\tempurl%
\url{https://doi.org/10.1007/978-3-319-46523-4_13}
\showDOI{\tempurl}


\bibitem[\protect\citeauthoryear{Du, Li, Torralba, Tenenbaum, and Mordatch}{Du et~al\mbox{.}}{2024}]%
        {du2024improving}
\bibfield{author}{\bibinfo{person}{Yilun Du}, \bibinfo{person}{Shuang Li}, \bibinfo{person}{Antonio Torralba}, \bibinfo{person}{Joshua~B. Tenenbaum}, {and} \bibinfo{person}{Igor Mordatch}.} \bibinfo{year}{2024}\natexlab{}.
\newblock \showarticletitle{Improving factuality and reasoning in language models through multiagent debate}. In \bibinfo{booktitle}{\emph{Proceedings of the 41st International Conference on Machine Learning}}. \bibinfo{publisher}{JMLR.org}, \bibinfo{address}{Vienna, Austria}, \bibinfo{pages}{11733--11763}.
\newblock


\bibitem[\protect\citeauthoryear{Hammar, Wallin, Karlberg, and H{\"a}lleberg}{Hammar et~al\mbox{.}}{2019}]%
        {hammar2019realestatecore}
\bibfield{author}{\bibinfo{person}{Karl Hammar}, \bibinfo{person}{Erik~Oskar Wallin}, \bibinfo{person}{Per Karlberg}, {and} \bibinfo{person}{David H{\"a}lleberg}.} \bibinfo{year}{2019}\natexlab{}.
\newblock \showarticletitle{The {RealEstateCore} Ontology}. In \bibinfo{booktitle}{\emph{The Semantic Web - ISWC 2019}}. \bibinfo{publisher}{Springer}, \bibinfo{address}{Auckland, New Zealand}, \bibinfo{pages}{130--145}.
\newblock
\urldef\tempurl%
\url{https://doi.org/10.1007/978-3-030-30796-7_9}
\showDOI{\tempurl}


\bibitem[\protect\citeauthoryear{Hertling and Paulheim}{Hertling and Paulheim}{2023}]%
        {hertling2023olala}
\bibfield{author}{\bibinfo{person}{Sven Hertling} {and} \bibinfo{person}{Heiko Paulheim}.} \bibinfo{year}{2023}\natexlab{}.
\newblock \showarticletitle{{OLaLa}: Ontology Matching with Large Language Models}. In \bibinfo{booktitle}{\emph{Proceedings of the 12th Knowledge Capture Conference 2023}}. \bibinfo{publisher}{ACM}, \bibinfo{address}{Pensacola, FL, USA}, \bibinfo{pages}{131--139}.
\newblock
\urldef\tempurl%
\url{https://doi.org/10.1145/3587259.3627571}
\showDOI{\tempurl}


\bibitem[\protect\citeauthoryear{Kalai, Nachum, Vempala, and Zhang}{Kalai et~al\mbox{.}}{2026}]%
        {kalai2026evaluating}
\bibfield{author}{\bibinfo{person}{Adam~Tauman Kalai}, \bibinfo{person}{Ofir Nachum}, \bibinfo{person}{Santosh~S Vempala}, {and} \bibinfo{person}{Edwin Zhang}.} \bibinfo{year}{2026}\natexlab{}.
\newblock \showarticletitle{Evaluating large language models for accuracy incentivizes hallucinations}.
\newblock \bibinfo{journal}{\emph{Nature}} (\bibinfo{year}{2026}), 29.
\newblock
\urldef\tempurl%
\url{https://doi.org/10.1038/s41586-026-10549-w}
\showDOI{\tempurl}


\bibitem[\protect\citeauthoryear{Li, Dragisic, Faria, Ivanova, Jiménez-Ruiz, Lambrix, and Pesquita}{Li et~al\mbox{.}}{2019}]%
        {li2019user}
\bibfield{author}{\bibinfo{person}{Huanyu Li}, \bibinfo{person}{Zlatan Dragisic}, \bibinfo{person}{Daniel Faria}, \bibinfo{person}{Valentina Ivanova}, \bibinfo{person}{Ernesto Jiménez-Ruiz}, \bibinfo{person}{Patrick Lambrix}, {and} \bibinfo{person}{Catia Pesquita}.} \bibinfo{year}{2019}\natexlab{}.
\newblock \showarticletitle{User validation in ontology alignment: functional assessment and impact}.
\newblock \bibinfo{journal}{\emph{The Knowledge Engineering Review}}  \bibinfo{volume}{34} (\bibinfo{year}{2019}), \bibinfo{pages}{e15}.
\newblock
\urldef\tempurl%
\url{https://doi.org/10.1017/S0269888919000080}
\showDOI{\tempurl}


\bibitem[\protect\citeauthoryear{McGowan}{McGowan}{2020}]%
        {john2020project}
\bibfield{author}{\bibinfo{person}{John~Jack McGowan}.} \bibinfo{year}{2020}\natexlab{}.
\newblock \showarticletitle{Project Haystack Data Standards}.
\newblock In \bibinfo{booktitle}{\emph{Energy and Analytics}}. \bibinfo{publisher}{River Publishers}, \bibinfo{address}{Aalborg, Denmark}, \bibinfo{pages}{237--243}.
\newblock
\urldef\tempurl%
\url{https://doi.org/10.1201/9781003151944-16}
\showDOI{\tempurl}


\bibitem[\protect\citeauthoryear{Nguyen, Barcelos, French, and Wu}{Nguyen et~al\mbox{.}}{2025}]%
        {nguyen2025kroma}
\bibfield{author}{\bibinfo{person}{Lam Nguyen}, \bibinfo{person}{Erika Barcelos}, \bibinfo{person}{Roger French}, {and} \bibinfo{person}{Yinghui Wu}.} \bibinfo{year}{2025}\natexlab{}.
\newblock \showarticletitle{{KROMA}: Ontology Matching with Knowledge Retrieval and Large Language Models}. In \bibinfo{booktitle}{\emph{The Semantic Web -- ISWC 2025}}. \bibinfo{publisher}{Springer}, \bibinfo{address}{Nara, Japan}, \bibinfo{pages}{629--649}.
\newblock
\urldef\tempurl%
\url{https://doi.org/10.1007/978-3-032-09527-5_34}
\showDOI{\tempurl}


\bibitem[\protect\citeauthoryear{{OAEI Community}}{{OAEI Community}}{nd}]%
        {oaei}
\bibfield{author}{\bibinfo{person}{{OAEI Community}}.} \bibinfo{year}{n.d.}\natexlab{}.
\newblock \bibinfo{title}{{Ontology Alignment Evaluation Initiative (OAEI)}}.
\newblock
\newblock
\urldef\tempurl%
\url{https://oaei.ontologymatching.org}
\showURL{%
Retrieved May 1, 2026 from \tempurl}


\bibitem[\protect\citeauthoryear{{OpenAI}}{{OpenAI}}{2022}]%
        {chatgpt}
\bibfield{author}{\bibinfo{person}{{OpenAI}}.} \bibinfo{year}{2022}\natexlab{}.
\newblock \bibinfo{title}{Introducing {ChatGPT}}.
\newblock
\newblock
\urldef\tempurl%
\url{https://openai.com/index/chatgpt/}
\showURL{%
Retrieved May 1, 2026 from \tempurl}


\bibitem[\protect\citeauthoryear{Ouyang, Wu, Jiang, Almeida, Wainwright, Mishkin, Zhang, Agarwal, Slama, Ray, Schulman, Hilton, Kelton, Miller, Simens, Askell, Welinder, Christiano, Leike, and Lowe}{Ouyang et~al\mbox{.}}{2022}]%
        {ouyang2022training}
\bibfield{author}{\bibinfo{person}{Long Ouyang}, \bibinfo{person}{Jeffrey Wu}, \bibinfo{person}{Xu Jiang}, \bibinfo{person}{Diogo Almeida}, \bibinfo{person}{Carroll Wainwright}, \bibinfo{person}{Pamela Mishkin}, \bibinfo{person}{Chong Zhang}, \bibinfo{person}{Sandhini Agarwal}, \bibinfo{person}{Katarina Slama}, \bibinfo{person}{Alex Ray}, \bibinfo{person}{John Schulman}, \bibinfo{person}{Jacob Hilton}, \bibinfo{person}{Fraser Kelton}, \bibinfo{person}{Luke Miller}, \bibinfo{person}{Maddie Simens}, \bibinfo{person}{Amanda Askell}, \bibinfo{person}{Peter Welinder}, \bibinfo{person}{Paul~F Christiano}, \bibinfo{person}{Jan Leike}, {and} \bibinfo{person}{Ryan Lowe}.} \bibinfo{year}{2022}\natexlab{}.
\newblock \showarticletitle{Training language models to follow instructions with human feedback}. In \bibinfo{booktitle}{\emph{Proceedings of the 36th International Conference on Neural Information Processing Systems}}, Vol.~\bibinfo{volume}{35}. \bibinfo{publisher}{Curran Associates, Inc.}, \bibinfo{address}{New Orleans, LA, USA}, \bibinfo{pages}{27730--27744}.
\newblock


\bibitem[\protect\citeauthoryear{Paulheim, Hertling, and Ritze}{Paulheim et~al\mbox{.}}{2013}]%
        {paulheim2013towards}
\bibfield{author}{\bibinfo{person}{Heiko Paulheim}, \bibinfo{person}{Sven Hertling}, {and} \bibinfo{person}{Dominique Ritze}.} \bibinfo{year}{2013}\natexlab{}.
\newblock \showarticletitle{Towards Evaluating Interactive Ontology Matching Tools}. In \bibinfo{booktitle}{\emph{The Semantic Web: Semantics and Big Data}}. \bibinfo{publisher}{Springer}, \bibinfo{address}{Montpellier, France}, \bibinfo{pages}{31--45}.
\newblock
\urldef\tempurl%
\url{https://doi.org/10.1007/978-3-642-38288-8_3}
\showDOI{\tempurl}


\bibitem[\protect\citeauthoryear{Qiang, Wang, and Taylor}{Qiang et~al\mbox{.}}{2024}]%
        {qiang2023agent}
\bibfield{author}{\bibinfo{person}{Zhangcheng Qiang}, \bibinfo{person}{Weiqing Wang}, {and} \bibinfo{person}{Kerry Taylor}.} \bibinfo{year}{2024}\natexlab{}.
\newblock \showarticletitle{{Agent-OM}: Leveraging {LLM} Agents for Ontology Matching}.
\newblock \bibinfo{journal}{\emph{Proceedings of the {VLDB} Endowment}} \bibinfo{volume}{18}, \bibinfo{number}{3} (\bibinfo{year}{2024}), \bibinfo{pages}{516--529}.
\newblock
\urldef\tempurl%
\url{https://doi.org/10.14778/3712221.3712222}
\showDOI{\tempurl}


\bibitem[\protect\citeauthoryear{Qiang, Wang, and Taylor}{Qiang et~al\mbox{.}}{2025}]%
        {qiang2025oaei}
\bibfield{author}{\bibinfo{person}{Zhangcheng Qiang}, \bibinfo{person}{Weiqing Wang}, {and} \bibinfo{person}{Kerry Taylor}.} \bibinfo{year}{2025}\natexlab{}.
\newblock \showarticletitle{{Agent-OM} Results for {OAEI} 2025}. In \bibinfo{booktitle}{\emph{The 20th International Workshop on Ontology Matching collocated with the 24th International Semantic Web Conference (ISWC 2025)}}, Vol.~\bibinfo{volume}{4144}. \bibinfo{publisher}{CEUR-WS.org}, \bibinfo{address}{Nara, Japan}, \bibinfo{pages}{202--210}.
\newblock


\bibitem[\protect\citeauthoryear{Rafailov, Sharma, Mitchell, Manning, Ermon, and Finn}{Rafailov et~al\mbox{.}}{2023}]%
        {rafailov2023direct}
\bibfield{author}{\bibinfo{person}{Rafael Rafailov}, \bibinfo{person}{Archit Sharma}, \bibinfo{person}{Eric Mitchell}, \bibinfo{person}{Christopher~D Manning}, \bibinfo{person}{Stefano Ermon}, {and} \bibinfo{person}{Chelsea Finn}.} \bibinfo{year}{2023}\natexlab{}.
\newblock \showarticletitle{Direct Preference Optimization: Your Language Model is Secretly A Reward Model}. In \bibinfo{booktitle}{\emph{Proceedings of the 37th International Conference on Neural Information Processing Systems}}, Vol.~\bibinfo{volume}{36}. \bibinfo{publisher}{Curran Associates Inc.}, \bibinfo{address}{New Orleans, LA, USA}, \bibinfo{pages}{53728--53741}.
\newblock


\bibitem[\protect\citeauthoryear{Sarasua, Simperl, and Noy}{Sarasua et~al\mbox{.}}{2012}]%
        {sarasua2012crowdmap}
\bibfield{author}{\bibinfo{person}{Cristina Sarasua}, \bibinfo{person}{Elena Simperl}, {and} \bibinfo{person}{Natalya~F. Noy}.} \bibinfo{year}{2012}\natexlab{}.
\newblock \showarticletitle{CrowdMap: Crowdsourcing Ontology Alignment with Microtasks}. In \bibinfo{booktitle}{\emph{The Semantic Web -- ISWC 2012}}. \bibinfo{publisher}{Springer}, \bibinfo{address}{Boston, MA, USA}, \bibinfo{pages}{525--541}.
\newblock
\urldef\tempurl%
\url{https://doi.org/10.1007/978-3-642-35176-1_33}
\showDOI{\tempurl}


\bibitem[\protect\citeauthoryear{Selgin}{Selgin}{1996}]%
        {selgin1996salvaging}
\bibfield{author}{\bibinfo{person}{George Selgin}.} \bibinfo{year}{1996}\natexlab{}.
\newblock \showarticletitle{Salvaging Gresham's Law: The good, the bad, and the illegal}.
\newblock \bibinfo{journal}{\emph{Journal of Money, Credit and Banking}} \bibinfo{volume}{28}, \bibinfo{number}{4} (\bibinfo{year}{1996}), \bibinfo{pages}{637--649}.
\newblock
\urldef\tempurl%
\url{https://doi.org/10.2307/2078075}
\showDOI{\tempurl}


\bibitem[\protect\citeauthoryear{Song, Chen, and Schmidt}{Song et~al\mbox{.}}{2026}]%
        {song2026genom}
\bibfield{author}{\bibinfo{person}{Yiping Song}, \bibinfo{person}{Jiaoyan Chen}, {and} \bibinfo{person}{Renate~A Schmidt}.} \bibinfo{year}{2026}\natexlab{}.
\newblock \showarticletitle{{GenOM}: ontology matching with description generation and large language models}.
\newblock \bibinfo{journal}{\emph{World Wide Web}} \bibinfo{volume}{29}, \bibinfo{number}{3} (\bibinfo{year}{2026}), \bibinfo{pages}{29}.
\newblock
\urldef\tempurl%
\url{https://doi.org/10.1007/s11280-026-01413-y}
\showDOI{\tempurl}


\bibitem[\protect\citeauthoryear{Sousa, Lima, and Trojahn}{Sousa et~al\mbox{.}}{2025}]%
        {sousa2025complex}
\bibfield{author}{\bibinfo{person}{Guilherme Sousa}, \bibinfo{person}{Rinaldo Lima}, {and} \bibinfo{person}{Cassia Trojahn}.} \bibinfo{year}{2025}\natexlab{}.
\newblock \bibinfo{title}{Complex Ontology Matching with Large Language Model Embeddings}.
\newblock
\newblock
\showeprint[arxiv]{2502.13619}~[cs.CL]
\urldef\tempurl%
\url{https://arxiv.org/abs/2502.13619}
\showURL{%
\tempurl}


\bibitem[\protect\citeauthoryear{Surowiecki}{Surowiecki}{2004}]%
        {surowiecki2004wisdom}
\bibfield{author}{\bibinfo{person}{James Surowiecki}.} \bibinfo{year}{2004}\natexlab{}.
\newblock \bibinfo{booktitle}{\emph{The Wisdom of Crowds}}.
\newblock \bibinfo{publisher}{Doubleday}, \bibinfo{address}{New York, NY, USA}.
\newblock


\bibitem[\protect\citeauthoryear{Sweller}{Sweller}{1988}]%
        {sweller1988cognitive}
\bibfield{author}{\bibinfo{person}{John Sweller}.} \bibinfo{year}{1988}\natexlab{}.
\newblock \showarticletitle{Cognitive load during problem solving: Effects on learning}.
\newblock \bibinfo{journal}{\emph{Cognitive Science}} \bibinfo{volume}{12}, \bibinfo{number}{2} (\bibinfo{year}{1988}), \bibinfo{pages}{257--285}.
\newblock
\urldef\tempurl%
\url{https://doi.org/10.1016/0364-0213(88)90023-7}
\showDOI{\tempurl}


\bibitem[\protect\citeauthoryear{Taboada, Martinez, Arideh, and Mosquera}{Taboada et~al\mbox{.}}{2025}]%
        {taboada2025ontology}
\bibfield{author}{\bibinfo{person}{Maria Taboada}, \bibinfo{person}{Diego Martinez}, \bibinfo{person}{Mohammed Arideh}, {and} \bibinfo{person}{Rosa Mosquera}.} \bibinfo{year}{2025}\natexlab{}.
\newblock \showarticletitle{Ontology matching with Large Language Models and prioritized depth-first search}.
\newblock \bibinfo{journal}{\emph{Information Fusion}}  \bibinfo{volume}{123} (\bibinfo{year}{2025}), \bibinfo{pages}{103254}.
\newblock
\urldef\tempurl%
\url{https://doi.org/10.1016/j.inffus.2025.103254}
\showDOI{\tempurl}


\bibitem[\protect\citeauthoryear{Tversky and Kahneman}{Tversky and Kahneman}{1974}]%
        {tversky1974judgment}
\bibfield{author}{\bibinfo{person}{Amos Tversky} {and} \bibinfo{person}{Daniel Kahneman}.} \bibinfo{year}{1974}\natexlab{}.
\newblock \showarticletitle{Judgment under Uncertainty: Heuristics and Biases}.
\newblock \bibinfo{journal}{\emph{Science}} \bibinfo{volume}{185}, \bibinfo{number}{4157} (\bibinfo{year}{1974}), \bibinfo{pages}{1124--1131}.
\newblock
\urldef\tempurl%
\url{https://doi.org/10.1126/science.185.4157.1124}
\showDOI{\tempurl}


\bibitem[\protect\citeauthoryear{Wei, Wang, Schuurmans, Bosma, Ichter, Xia, Chi, Le, and Zhou}{Wei et~al\mbox{.}}{2022}]%
        {wei2023chainofthought}
\bibfield{author}{\bibinfo{person}{Jason Wei}, \bibinfo{person}{Xuezhi Wang}, \bibinfo{person}{Dale Schuurmans}, \bibinfo{person}{Maarten Bosma}, \bibinfo{person}{Brian Ichter}, \bibinfo{person}{Fei Xia}, \bibinfo{person}{Ed~H. Chi}, \bibinfo{person}{Quoc~V. Le}, {and} \bibinfo{person}{Denny Zhou}.} \bibinfo{year}{2022}\natexlab{}.
\newblock \showarticletitle{Chain-of-Thought Prompting Elicits Reasoning in Large Language Models}. In \bibinfo{booktitle}{\emph{Proceedings of the 36th Annual Conference on Neural Information Processing Systems}}, Vol.~\bibinfo{volume}{35}. \bibinfo{publisher}{Curran Associates, Inc.}, \bibinfo{address}{New Orleans, LA, USA}, \bibinfo{pages}{24824--24837}.
\newblock


\bibitem[\protect\citeauthoryear{Zhang, Dong, Zhang, Payne, and Zhang}{Zhang et~al\mbox{.}}{2024}]%
        {zhang2024large}
\bibfield{author}{\bibinfo{person}{Shiyao Zhang}, \bibinfo{person}{Yuji Dong}, \bibinfo{person}{Yichuan Zhang}, \bibinfo{person}{Terry~R. Payne}, {and} \bibinfo{person}{Jie Zhang}.} \bibinfo{year}{2024}\natexlab{}.
\newblock \showarticletitle{Large Language Model Assisted Multi-Agent Dialogue for Ontology Alignment}. In \bibinfo{booktitle}{\emph{Proceedings of the 2024 International Conference on Autonomous Agents and Multiagent Systems}}. \bibinfo{publisher}{IFAAMAS}, \bibinfo{address}{Auckland, New Zealand}, \bibinfo{pages}{2594--2596}.
\newblock


\end{thebibliography}

\end{document}